# Coherent Driving of a Single Nitrogen Vacancy Center by a Resonant Magnetic Tunnel Junction

Gerald Q. Yan[1,2,+], Nathan McLaughlin[2,+], Tatsuya Yamamoto[3], Senlei Li[1], Takayuki Nozaki[3], Shinji Yuasa[3], Chunhui Rita Du[1,2,*], and Hailong Wang[1,*]

[1]School of Physics, Georgia Institute of Technology, Atlanta, Georgia 30332, USA
[2]Department of Physics, University of California, San Diego, La Jolla, California 92093, USA
[3]National Institute of Advanced Industrial Science and Technology (AIST), Research Center for Emerging Computing Technologies, Tsukuba, Ibaraki, 305-8568, Japan

[*]Corresponding authors: cdu71@gatech.edu; hwang3021@gatech.edu
[+]These authors contributed equally.

**Abstract**: Nitrogen-vacancy (NV) centers, atomic spin defects in diamond, represent an active contender for advancing transformative quantum information science (QIS) and innovations. One of the major challenges for designing NV-based hybrid systems for QIS applications results from the difficulty of realizing local control of individual NV spin qubits in a scalable and energy-efficient way. To address this bottleneck, we introduce magnetic tunnel junction (MTJ) devices to establish coherent driving of an NV center by a resonant MTJ with voltage controlled magnetic anisotropy. We show that the oscillating magnetic stray field produced by a resonant micromagnet can be utilized to effectively modify and drive NV spin rotations when the NV frequency matches the corresponding resonance conditions of the MTJ. Our results present a new pathway to achieve all-electric control of an NV spin qubit with reduced power consumption and improved solid-state scalability for implementing cutting-edge QIS technological applications.

Quantum sensing, computing, and communication are rapidly developing fields with the potential to revolutionize a wide range of emerging technological applications.[1–4] Various solid-state qubit systems are being explored under this context, such as trapped ions,[5] superconducting qubits,[6,7] and semiconductor quantum dots,[8] each presenting their own advantages and limitations. Nitrogen-vacancy (NV) centers,[9,10] optically active spin defects in diamond, represent another promising candidate for these purposes. NV defects possess atomically small dimensions and excellent quantum coherence, making them ideal sensors for probing electromagnetic fields with ultra-high sensitivity and spatial resolution.[10–12] Due to their discrete spin energy levels, NV centers naturally couple with photons, enabling photon-mediated spin manipulation and long-range entanglement protocols for quantum communications and networking applications.[4,13,14] Furthermore, by leveraging the long coherence times of the surrounding 13C and 14N nuclear spins, researchers have realized proof-of-concept demonstrations of quantum computing using single NV centers.[14–21] More recent successes in this direction include a ten-qubit spin memory[22] as well as the implementation of Grover's search algorithm.[23]

Nevertheless, one of the main difficulties hindering the development of NV-based quantum computing and relevant quantum applications is the local control of individual NV qubits in a scalable and energy efficient way.[18,19,24] At the present technological levels, NV spin states are typically manipulated by the spatially dispersed Oersted fields generated by radiofrequency (RF) currents flowing through an on-chip stripline,[25] leading to substantial "cross-talk" between neighboring NV centers, and reduced qubit density and scalability for practical applications. To address these longstanding issues, here we report electrical control of a single NV center qubit in a NV-magnetic tunnel junction (MTJ)[26,27]-based hybrid system. We show that the oscillating magnetic stray fields produced by the resonant MTJ can be utilized to effectively modify and drive coherent NV spin rotations at resonant field conditions.

We first describe the detailed device structure and measurement platform for NV control experiments as shown in Fig. 1a. The single NV center is introduced via a diamond cantilever, which is attached to the end of an atomic force microscopy (AFM) tuning fork allowing for nanoscale positioning and scanning measurements.[28,29] An overhead optical image and a diagram of the MTJ device exhibiting voltage-controlled magnetic anisotropy (VCMA) used in the study are shown in Fig. 1b. The MTJ stack is comprised of (from bottom to top): Ta(5 nm)/Co$_{40}$Fe$_{40}$B$_{20}$(1 nm)/MgO(2 nm)/Co$_{56}$Fe$_{24}$B$_{20}$(5 nm)/Ru(7 nm)/Cr(5 nm)/Au(50 nm) and is fabricated into an oval shape with a length of 6 μm and a width of 2 μm.[17] The MTJs were prepared using standard photo-lithography, dry etching, and sputtering processes, the details of which were reported previously.[26] The upper Co$_{56}$Fe$_{24}$B$_{20}$ layer exhibits spontaneous in-plane magnetization and serves as the magnetic reference layer, whilst the lower Co$_{40}$Fe$_{40}$B$_{20}$ serves as the free layer which under zero external magnetic field is perpendicularly magnetized due to a weak out-of-plane anisotropy.[17] Au contacts were patterned on the top and bottom of the MTJ for electrical connections.

Dipole coupling between the NV center and the MTJ takes advantage of the local oscillating magnetic stray field generated by the VCMA-driven magnetic resonance.[26,27] The magnetic easy axis of the free Co$_{40}$Fe$_{40}$B$_{20}$ layer can be switched between the out-of-plane and in-plane directions depending on the sign of the electrical voltage across the MTJ.[17,26] In an intuitive picture, a microwave voltage applied to the MTJ will drive gigahertz scale oscillations, also referred to as magnetic resonance, of the Co$_{40}$Fe$_{40}$B$_{20}$ free layer magnetization under appropriate external magnetic fields and microwave frequencies.[26,27] For the NV and VCMA-induced magnetic resonance measurements presented in this work, an external magnetic field $B_{ext}$ is applied at an angle of 54 degrees relative to the out-of-plane direction, along the NV spin axis and also in an ideal orientation to maximize the magnetic resonance response of the MTJ.[26] The in-plane projection of $B_{ext}$ lies along the long-axis of the MTJ.

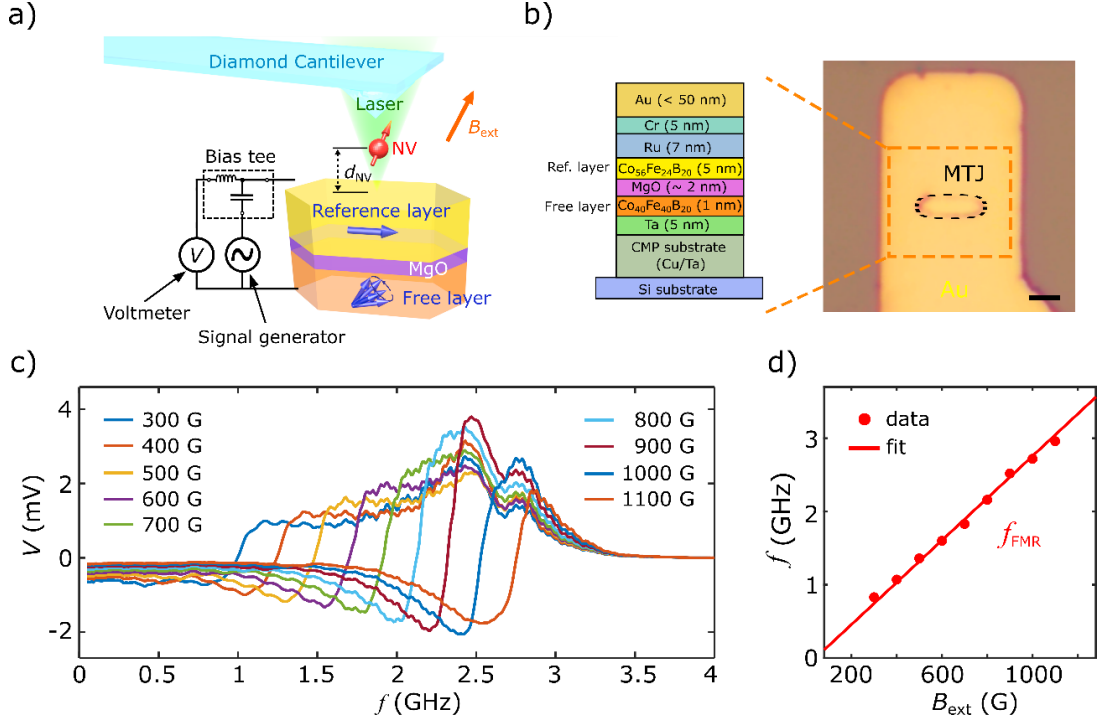

**Figure 1.** (a) Schematic illustration of a hybrid system consisting of a single NV center contained in scanning diamond cantilever and a MTJ device. The NV center is suspended a distance $d_{NV}$ from the top surface of the MTJ and is optically addressed using a confocal microscope. Electrical excitation and detection of VCMA-induced magnetic resonance utilizes a standard homodyne detection circuit. (b) Optical microscope image of an MTJ device which is outlined by the dashed black lines. The scale bar is 2 μm. The inset shows a cross-sectional schematic of the MTJ device. (c) Homodyne detection DC voltage $V$ measured as a function of the applied frequency $f$ under different external magnetic fields. (d) Fitted resonant frequencies (red points) of the MTJ device as a function of the applied external magnetic field. The data fits well with the expected modified Kittel equation (red solid line).

To electrically excite and detect the magnetic resonance of the MTJ, the homodyne detection circuit shown in Fig. 1a was used.[26] An RF voltage (−2 dBm) is applied to the MTJ through the RF port of a bias tee. The oscillating magnetization of the $Co_{40}Fe_{40}B_{20}$ free layer leads to a time dependent tunneling magnetoresistance (TMR), which couples with the RF current tunneling through the MgO barrier to produce a DC voltage $V$, which is subsequently detected through the DC port of the bias tee.[26] Figure 1c shows the measured homodyne DC signals as a function of the applied microwave frequency $f$ for various values of external magnetic field $B_{ext}$. By fitting each ferromagnetic resonance (FMR) spectrum with a sum of a Lorentzian and an anti-Lorentzian function,[17,26] the resonant frequency $f_{FMR}$ can be extracted. The obtained $f_{FMR}$ are plotted in Fig. 1d, showing the characteristic field and frequency dependence according to the modified Kittel formula:[17,26]

$$f_{FMR} = \frac{\gamma}{2\pi}\sqrt{\{B_{ext} - B_{d,eff}\cos^2\theta_H + B_{d,\text{in-plane}}\}\{B_{ext} - B_{d,eff}\cos(2\theta_H)\}} \quad (1)$$

where $\gamma/2\pi = 2.9 \times 10^6$ (s$^{-1}$·G$^{-1}$) is the gyromagnetic ratio for the magnetic free layer, $B_{d,eff} = -135$ G is the effective demagnetization field along the out-of-plane direction, $\theta_H = 54$ degrees, and $B_{d,in\text{-}plane} = -88$ G is the difference between in-plane demagnetization fields along the short and long-axis directions of the MTJ.

We next use scanning NV magnetometry to examine the static magnetic stray field environment of the MTJ sample. A diamond cantilever containing a single NV electron spin is positioned above the sample (Fig. 2a).[28,29] Figure 2b displays an AFM image of the topography of the MTJ device. The scanning AFM capabilities allow for nanoscale control of the vertical NV-to-sample distance $d_{NV}$, which ultimately dictates the spatial resolution of the NV magnetometry measurements and the strength of the static and microwave MTJ fields felt by the NV center.[11] For the NV measurements presented in the current work, $d_{NV}$ is fixed at 250 nm unless stated otherwise. The NV center is an optically addressable spin defect consisting of a substitutional nitrogen atom and an adjacent carbon atom vacancy in a diamond crystal lattice.[9,10] The negatively charged NV state has an $S = 1$ electron spin and serves as a "three-level" quantum system.[9] NV magnetometry allows for accurate magnetic field sensing at the nanoscale. The magnetic moment of the NV center is a powerful magnetic analyzer, due to the Zeeman effect, the energy levels of the NV center experience a splitting by an amount that depends on the applied external magnetic field.[10] The NV spin states can then be optically addressed using spin-dependent photoluminescence (PL) measurements. By probing the energy splitting between the two PL peaks, the magnitude of local static magnetic field parallel to the NV spin axis can be obtained. By scanning the NV center and performing optically detected magnetic resonance (ODMR) measurements at each position above the MTJ, a spatial map of the stray field $B_s$ emanating from the MTJ can be obtained as shown in Fig. 2c. It is evident that the magnetic stray field mainly emerges from the two opposing edges along the long axis of the MTJ, as is expected for a largely uniform magnetic distribution along the easy axis. Due to the dipolar nature, the emanating magnetic stray field changes sign across each of the two sample edges. The obtained dipole stray field pattern indicates that the dominant contribution is due to the in-plane magnetized reference layer which has a larger thickness and stronger magnetization than the free layer. We also perform micromagnetic simulations of the stray field $B_s$ distribution of the MTJ sample as shown in Fig. 2d, which is in excellent agreement with our scanning NV results.

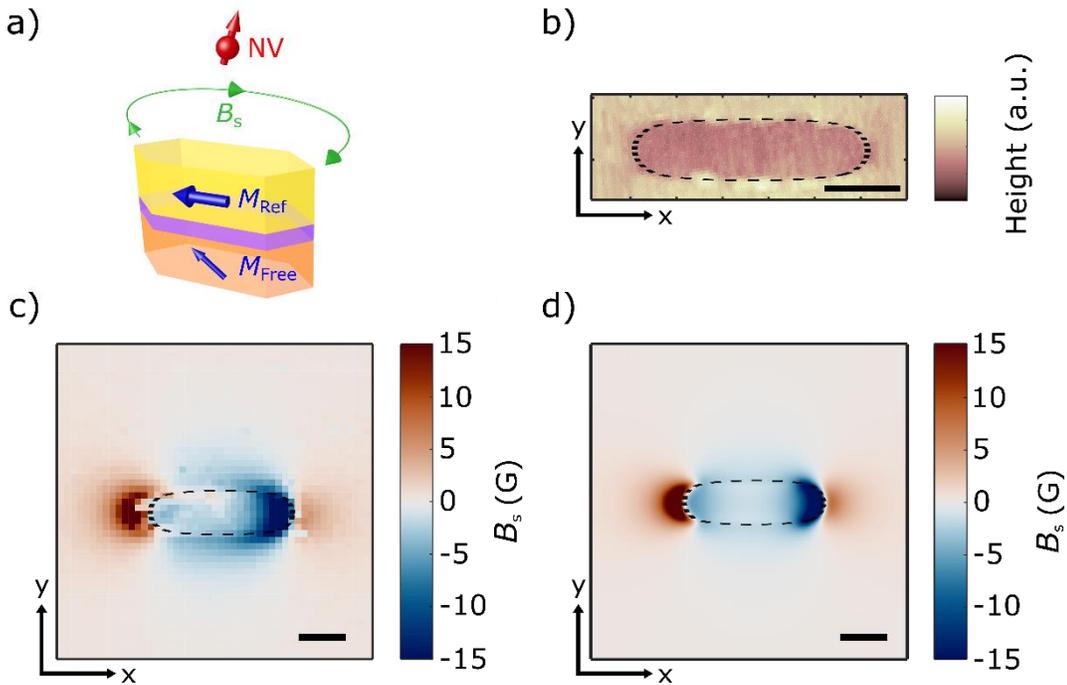

**Figure 2.** (a) A diagram of the static stray field $B_s$ at the NV site due to magnetic layers of the MTJ. (b) AFM imaging of the topography of an MTJ device. (c) 2D magnetic static stray map of the MTJ device. An external magnetic field of 456 G is applied along the NV axis in this measurement. (d) Simulated magnetic stray field map of the MTJ device. The black dashed lines outline the edges of the MTJ, and all scale bars are 2 μm.

We now present resonant MTJ-assisted coherent control of the single NV qubit. Figure 3a shows the schematic of the measurement system, where RF currents (15 dBm) driven through an external Au microwave wire produce an oscillating Oersted field $B_{MW}$ for exciting the NV spin states. Under the appropriate field conditions, the external microwave currents can also excite magnetic resonance of the MTJ free layer,[17,26] producing an additional oscillating magnetic stray field $B_{RES}$. When $B_{MW}$ and/or $B_{RES}$ match the NV ESR frequencies $f_\pm$, coherent NV spin rotations between the $m_s = 0$ and $m_s = \pm 1$ spin state(s) will be excited, which are referred to as NV Rabi oscillations.[1,9,10] The top panel of Fig. 3b shows the measurement protocol for NV Rabi oscillation using microwave currents flowing through the Au wire. A green laser pulse is first used to initialize the NV qubit to the $m_s = 0$ state. Next, an RF current pulse with variable duration $t$ at the NV ESR frequency $f_-$ is applied to drive the $m_s = 0 \leftrightarrow -1$ NV spin transition, whilst simultaneously exciting FMR of the magnetic free layer of the MTJ under the corresponding resonance condition.[17] A second green laser pulse is then applied to optically read out the NV spin state. As the microwave pulse duration is swept, the NV occupation probabilities will oscillate between the $m_s = 0$ and $m_s = -1$ states, leading to time dependent variations of the NV PL.[30] The bottom panel of Fig. 3b highlights the intuitive NV Rabi oscillation picture discussed above, wherein the component of total effective microwave magnetic field $B_{TOT\perp}$ (including $B_{MW}$ and $B_{RES}$) that is transverse to the NV spin axis will drive coherent NV rotations around the Bloch sphere.

Next, we present NV Rabi oscillation measurements performed under different NV ESR frequencies to investigate the corresponding contribution from the resonant MTJ for coherent NV spin control. Figure 3c plots the theoretically calculated FMR and parametric resonance curves of the magnetic free layer of the MTJ against the NV ESR frequency $f_-$. The FMR curve is found to intersect with the NV ESR curve at a field of ~448 G and a frequency of ~1.62 GHz. At sufficiently large input microwave driving powers, parametric resonance can also be excited in the MTJ free layer, leading to higher-order collective spin waves with precession axis of the magnetic moments parallel to the external microwave magnetic field direction.[31,32] The parametric resonance has a characteristic threshold frequency value of twice of the magnon band minimum[33] and its dispersion curve is expected to intersect with $f_-$ at a magnetic field of ~260 G and a frequency of ~2.14 GHz. Figure 3d presents the measured NV Rabi oscillation spectra measured under three different external magnetic fields $B_{ext}$ of 448 G, 259 G, and 176 G, corresponding to the crossing NV frequency with the MTJ FMR curve, MTJ parametric curve, and an off-resonant point, respectively. At a field of 176 G, NV Rabi oscillations are observed with a Rabi frequency $f_{Rabi}$ of ~5.7 MHz. It is evident that the NV center shows enhanced Rabi oscillation rates when $f_-$ crosses with the FMR and parametric dispersion curves. At a field of 448 G, the measured NV PL spectrum exhibits accelerated oscillatory behavior with an increased $f_{Rabi}$ of ~14.4 MHz. At the crossing point with the parametric curve (~259 G), moderately enhanced Rabi oscillations are also observed with $f_{Rabi}$ of ~7.7 MHz. The enhancement of the NV coherent spin rotation rate results from the extra oscillating magnetic stray field $B_{RES}$ generated by the resonant MTJ, which amplifies the effective microwave magnetic field experienced by the NV center. Figure 3e summarizes the measured NV Rabi frequencies as a function of external magnetic field $B_{ext}$ along the corresponding NV ESR curve $f_-$. One can see that at the FMR and parametric conditions, the resonant MTJ strongly couples to the NV center, driving faster NV spin rotations. It is instructive to note that, due to the coherent nature of the magnetization dynamics, the FMR-driven magnetization precession produces a larger oscillating magnetic field than that generated by the parametric resonance, thus, it is reasonable to expect a higher NV Rabi oscillation frequency at the corresponding resonance condition. When detuned from the resonance frequencies of the MTJ, the measured NV Rabi frequencies gradually drop to the intrinsic values, indicating a negligible dipole coupling with the MTJ. At these points, the NV spin oscillations are driven solely by the external microwave currents flowing through the Au wire.

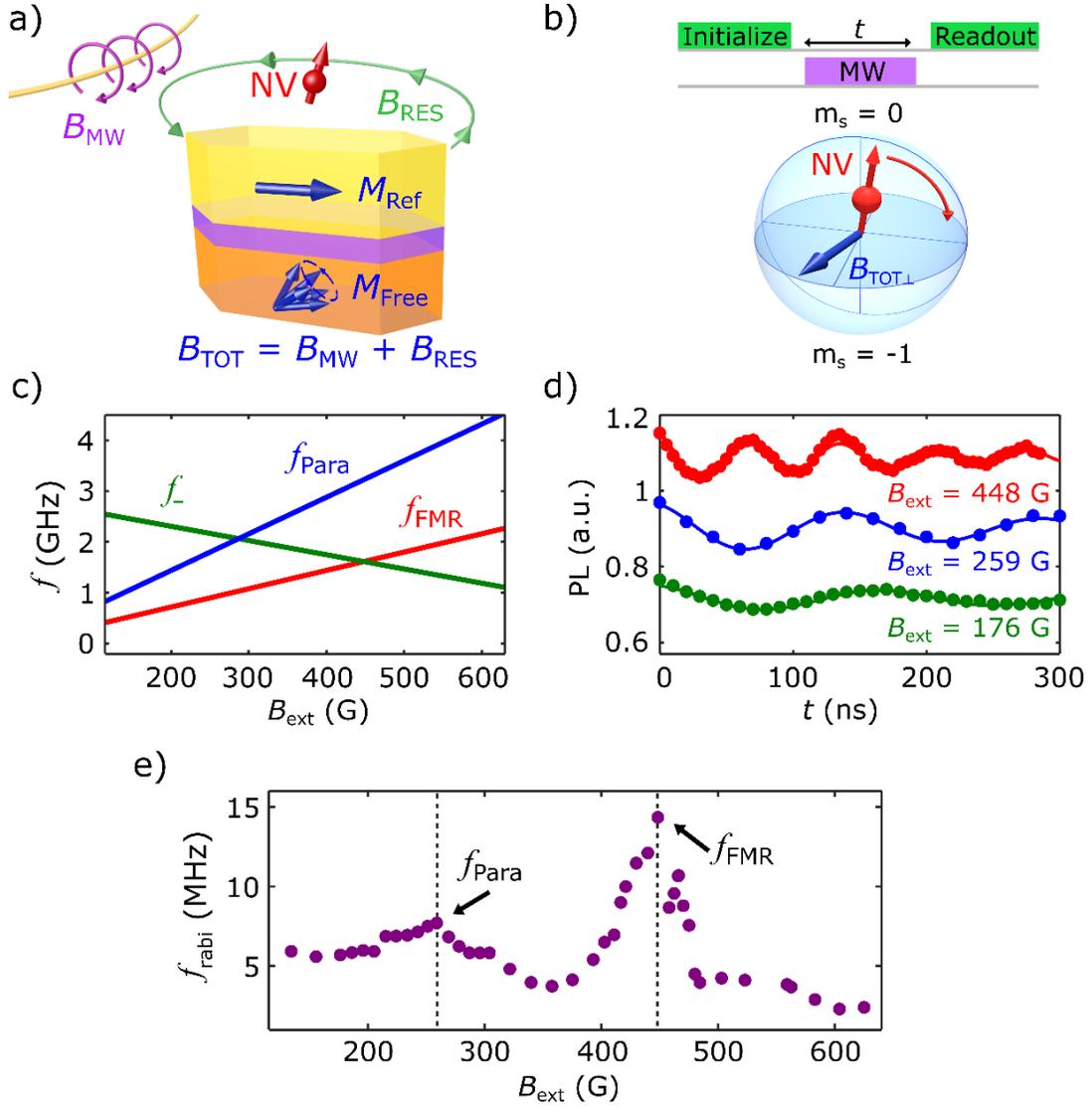

**Figure 3.** (a) Schematic of external Au microwave wire driving of NV spin rotations and MTJ resonances. RF currents (15 dBm) are delivered through the microwave wire to generate spatially dispersed Oersted fields. At the NV site, the total field $B_{TOT}$ includes contributions from the Oersted field $B_{MW}$ and the oscillating magnetic stray field $B_{RES}$ due to the resonant dynamics of the MTJ free layer. (b) Top panel: Optical and RF current pulse (applied in the external microwave wire) sequence to drive NV Rabi oscillations. Bottom panel: Schematic of NV Rabi oscillations between the $m_s = 0$ and $m_s = -1$ state on the Bloch sphere. (c) FMR (red) and parametric (blue) resonance frequency of the MTJ plotted as a function of the applied frequency $f$ and external magnetic field $B_{ext}$. The NV ESR curve $f_-$ corresponding to the $m_s = 0 \leftrightarrow -1$ spin transition is plotted in green. (d) NV Rabi oscillation spectra measured under different external magnetic fields corresponding to the FMR crossing point, parametric frequency crossing point, and a low-field off-resonant point, at external magnetic fields of 448 G, 259 G, and 176 G, respectively. (e) Measured NV Rabi frequencies $f_{Rabi}$ as a function of external magnetic field $B_{ext}$.

Lastly, we demonstrate resonant MTJ-driven coherent control of the NV spin qubit without applying external RF currents through the Au stripline as illustrated in Fig. 4a. The top panel of Fig. 4b shows the measurement protocol used for this study. After initializing the NV spin using a green laser pulse, an RF voltage pulse (−2 dBm) at the NV ESR frequency $f_-$ is applied across the MTJ.[17,26] When $f_-$ matches the FMR and parametric resonant frequencies ($f_{FMR}$ and $f_{para}$), oscillating magnetic stray fields will be generated by the MTJ to manipulate the NV spin states, which are then optically read out by a second laser pulse. The bottom panel of Fig. 4b presents the measured NV Rabi oscillation spectra measured at $f_-$ = 1.45 GHz, 2.22 GHz, and 1.31 GHz. When the NV ESR frequency intersects with the FMR curve, we see strongly enhanced Rabi oscillations with a Rabi frequency $f_{Rabi}$ of ~9.4 MHz, indicating a strong driving of the NV center by the resonant MTJ. At the parametric crossing frequency ($f_- = f_{para}$), moderate Rabi oscillations with a Rabi frequency of ~6.6 MHz are observed. A representative Rabi PL spectrum taken at an off-resonant point ($f_-$ = 1.31 GHz) exhibits diminished Rabi oscillation frequency, indicative of the suppressed NV-MTJ coupling as the driving frequency $f_-$ is detuned from the crossing frequencies. Figure 4c summarizes the measured $f_{Rabi}$ as a function of the frequency of the input RF voltage pulse. Clear peaks are observed at the NV ESR intersection frequencies with the FMR and parametric curves of the MTJ. The dramatic enhancement results from the phase synchronization between the NV qubit and the oscillating magnetization dynamics of the resonant MTJ. Fitting the peaks of the Rabi spectra with the equation $PL(t) = Ae^{-\frac{t}{\tau}} + B$, where $A$ and $B$ are fitting parameters, the NV coherence times are obtained to be 0.7 μs, 0.6 μs, and 1.2 μs at the FMR crossing, parametric crossing, and off-resonant frequencies, respectively.

To investigate the spatial distribution of the oscillating magnetic field generated by the resonant MTJ, we scan the diamond cantilever across the short axis of the sample whilst Rabi measurements are taken, at an external magnetic field of ~290 G. From the measured Rabi frequencies, the component of the oscillating magnetic stray field transverse to the NV axis $B_\perp$ can be extracted. Figure 4d shows a one-dimensional scan of the extracted transverse microwave field $B_\perp$ across the width of the MTJ sample. $B_\perp$ is found to show a finite value within the physical confines of the MTJ, before quickly decaying to a vanishingly small value at positions beyond the width of the sample. Theoretical simulations are in qualitative agreement with our experimental results, confirming the "localized" nature of the oscillating magnetic stray field produced by the resonant MTJ. We further evaluate the spatial dependence of $B_\perp$ as a function of $d_{NV}$ as presented in Fig. 4e. The NV center is initially situated at the center of the MTJ at $d_{NV}$ = 100 nm, with each subsequent Rabi measurement taken at increasing NV standoff distances. A monotonic decrease of $B_\perp$ is observed with increasing $d_{NV}$. The MTJ oscillating stray field remains capable of driving coherent NV Rabi oscillations for a larger $d_{NV}$ up to the micrometer scale, showing the potential for developing functional quantum spintronic devices based on NV centers and resonant microwave magnetic devices.[15,31,34,35]

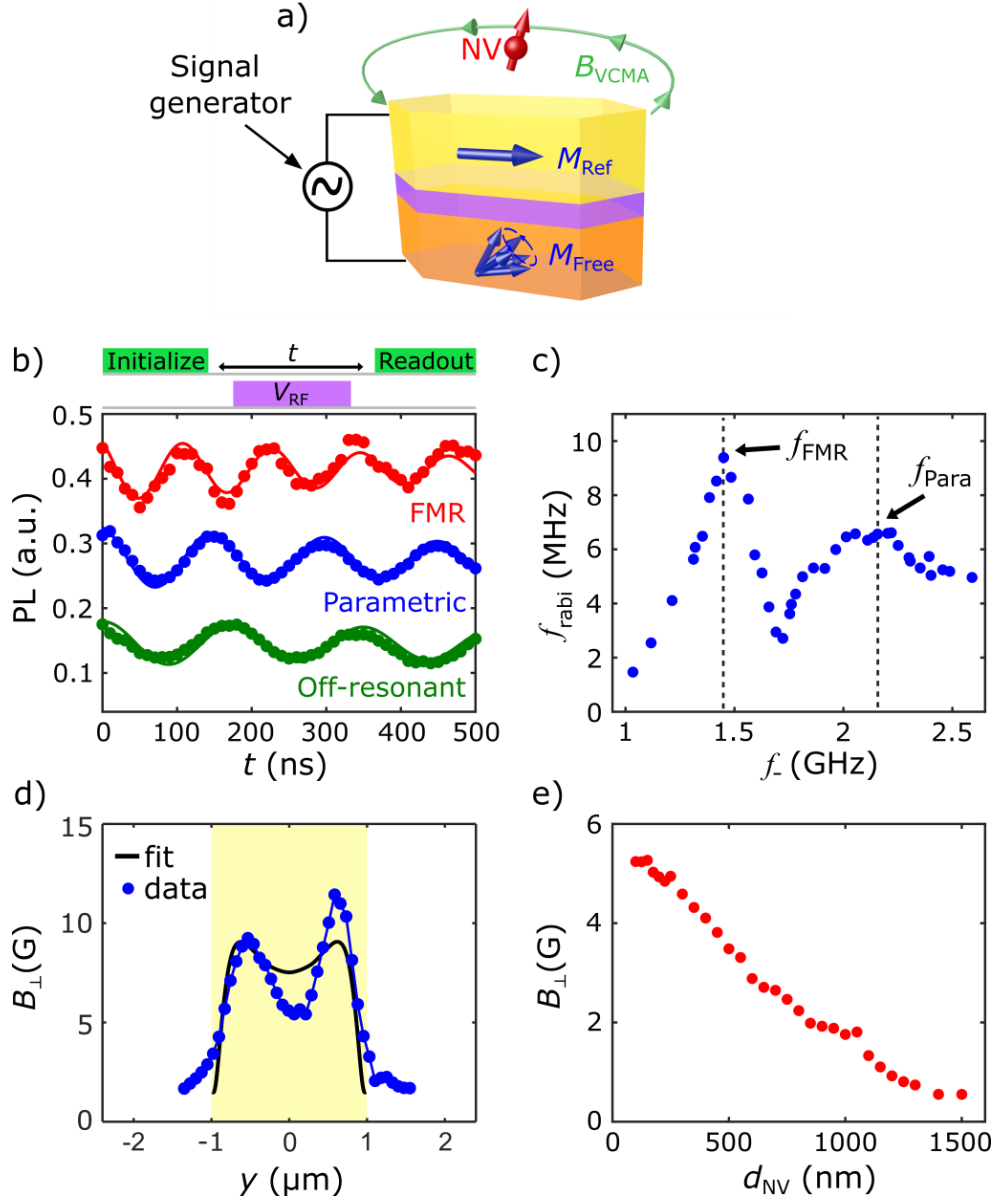

**Figure 4.** (a) Schematic of VCMA-induced MTJ resonances and coherent NV spin rotations. RF voltage pulses (−2 dBm) at the NV ESR frequency $f_-$ are applied across the MTJ. At resonance conditions, VCMA-driven FMR will produce oscillating magnetic stray fields which couple to the proximal NV center. (b) Top panel: Optical and RF voltage pulse sequence to drive NV Rabi oscillations. Bottom panel: RF voltage pulse time $t$ dependent NV PL spectra under different external magnetic fields corresponding to the FMR crossing point, parametric frequency crossing point, and a low-field off-resonant point. (c) NV Rabi frequency $f_{Rabi}$ as a function of NV ESR frequency $f_-$ for resonant MTJ-driven Rabi oscillation measurements. (d) Spatial dependence of transverse magnetic stray field $B_\perp$ measured across the short axis ($y$-axis) of the MTJ device. The black solid line plots the calculated one-dimensional $B_\perp$ along the same linecut, and the yellow shaded area represents the lateral width of the MTJ device along the short axis ($y$-axis). (e) Measured transverse magnetic stray field $B_\perp$ as a function of the NV-MTJ distance $d_{NV}$.

In summary, we have demonstrated coherent driving of a single NV center by a resonant MTJ device in a hybrid spintronic system. We show that a resonant MTJ can effectively amplify the magnitude of the local microwave fields experienced by a proximal NV center, leading to significant enhancement of its Rabi oscillation frequencies. Taking advantage of the VCMA of the MTJ, the oscillating magnetic stray fields generated by the resonant micromagnet (free layer) can be further utilized to achieve all-electric control of an NV spin qubit at reduced microwave power consumption. It is worth mentioning that the emanating magnetic stray field spatially decays more rapidly than the Oersted field generated by microwave currents. The "localized" nature of this NV qubit control scheme based on characteristic magnetic dipole interactions offers a new route for designing NV-based, high-density, and scalable quantum systems with excellent coherence performance for cutting-edge sensing, computing, and quantum information science applications.[15,34]

**Author contributions:**
G. Q. Y. and. N. M. performed the measurements and analyzed the data with help from S. L. T. Y., T. N., and S. Y. prepared and provided the magnetic tunnel junction devices. C. H. R. D. and H. W. supervised this project. All authors contributed to writing the manuscript.
**Notes:** The authors declare no financial interest.

**Acknowledgements**:
G. Q. Y. and C. H. R. D. were supported by U. S. National Science Foundation (NSF) under award no. ECCS-2029558. N. M. acknowledged the support from the Air Force Office of Scientific Research under award no. FA9550-20-1-0319. C. H. R. D. also acknowledges the support from the Office of Naval Research (ONR) under grant no. N00014-23-1-2146.

**References**

(1) Degen, C. L.; Reinhard, F.; Cappellaro, P. Quantum Sensing. *Rev. Mod. Phys.* **2017**, *89*, 035002.
(2) Popkin, G. Quest for Qubits. *Science.* **2016**, 354, 1090-1093.
(3) Kimble, H. J. The Quantum Internet. *Nature* **2008**, *453*, 1023–1030.
(4) Hensen, B.; Bernien, H.; Dréau, A. E.; Reiserer, A.; Kalb, N.; Blok, M. S.; Ruitenberg, J.; Vermeulen, R. F. L.; Schouten, R. N.; Abellán, C.; Amaya, W.; Pruneri, V.; Mitchell, M. W.; Markham, M.; Twitchen, D. J.; Elkouss, D.; Wehner, S.; Taminiau, T. H.; Hanson, R. Loophole-Free Bell Inequality Violation Using Electron Spins Separated by 1.3 Kilometres. *Nature* **2015**, *526*, 682–686.
(5) Cirac, J. I.; Zoller, P. Quantum Computations with Cold Trapped Ions. *Phys. Rev. Lett.* **1995**, *74*, 4091–4094.
(6) Devoret, M. H.; Schoelkopf, R. J. Superconducting Circuits for Quantum Information: An Outlook. *Science* **2013**, *339*, 1169–1174.
(7) Arute, F.; Arya, K.; Babbush, R.; Bacon, D.; Bardin, J. C.; Barends, R.; Biswas, R.; Boixo, S.; Brandao, F. G. S. L.; Buell, D. A.; Burkett, B.; Chen, Y.; Chen, Z.; Chiaro, B.; Collins, R.; Courtney, W.; Dunsworth, A.; Farhi, E.; Foxen, B.; Fowler, A.; Gidney, C.; Giustina, M.; Graff, R.; Guerin, K.; Habegger, S.; Harrigan, M. P.; Hartmann, M. J.; Ho, A.; Hoffmann, M.; Huang, T.; Humble, T. S.; Isakov, S. V.; Jeffrey, E.; Jiang, Z.; Kafri, D.; Kechedzhi, K.; Kelly, J.; Klimov, P. V.; Knysh, S.; Korotkov, A.; Kostritsa, F.; Landhuis, D.; Lindmark, M.; Lucero, E.; Lyakh, D.; Mandrà, S.; McClean, J. R.; McEwen, M.; Megrant, A.; Mi, X.; Michielsen, K.; Mohseni, M.; Mutus, J.; Naaman, O.; Neeley, M.; Neill, C.; Niu, M. Y.; Ostby, E.; Petukhov, A.; Platt, J. C.; Quintana, C.; Rieffel, E. G.; Roushan, P.; Rubin, N. C.; Sank, D.; Satzinger, K. J.; Smelyanskiy, V.; Sung, K. J.; Trevithick, M. D.; Vainsencher, A.; Villalonga, B.; White, T.; Yao, Z. J.; Yeh, P.; Zalcman, A.; Neven, H.; Martinis, J. M. Quantum Supremacy Using a Programmable Superconducting Processor. *Nature* **2019**, *574*, 505–510.
(8) Burkard, G.; Ladd, T. D.; Pan, A.; Nichol, J. M.; Petta, J. R. Semiconductor Spin Qubits. *Rev. Mod. Phys.* **2023**, *95*, 025003.
(9) Doherty, M. W.; Manson, N. B.; Delaney, P.; Jelezko, F.; Wrachtrup, J.; Hollenberg, L. C. L. The Nitrogen-Vacancy Colour Centre in Diamond. *Phys. Rep.* **2013**, *528*, 1–45.
(10) Rondin, L.; Tetienne, J.-P.; Hingant, T.; Roch, J.-F.; Maletinsky, P.; Jacques, V. Magnetometry with Nitrogen-Vacancy Defects in Diamond. *Rep. Prog. Phys.* **2014**, *77*, 056503.
(11) Casola, F.; Van Der Sar, T.; Yacoby, A. Probing Condensed Matter Physics with Magnetometry Based on Nitrogen-Vacancy Centres in Diamond. *Nat. Rev. Mater.* **2018**, *3*, 17088.
(12) Dolde, F.; Fedder, H.; Doherty, M. W.; Nöbauer, T.; Rempp, F.; Balasubramanian, G.; Wolf, T.; Reinhard, F.; Hollenberg, L. C. L.; Jelezko, F.; Wrachtrup, J. Electric-Field Sensing Using Single Diamond Spins. *Nat. Phys.* **2011**, *7*, 459–463.
(13) Ruf, M.; Wan, N. H.; Choi, H.; Englund, D.; Hanson, R. Quantum Networks Based on Color Centers in Diamond. *J. Appl. Phys.* **2021**, *130*, 070901.
(14) Childress, L.; Hanson, R. Diamond NV Centers for Quantum Computing and Quantum Networks. *MRS Bull.* **2013**, *38*, 134–138.
(15) Awschalom, D. D.; Du, C. R.; He, R.; Heremans, F. J.; Hoffmann, A.; Hou, J.; Kurebayashi, H.; Li, Y.; Liu, L.; Novosad, V.; Sklenar, J.; Sullivan, S. E.; Sun, D.; Tang, H.; Tyberkevych, V.; Trevillian, C.; Tsen, A. W.; Weiss, L. R.; Zhang, W.; Zhang, X.; Zhao, L.; Zollitsch, Ch.

W. Quantum Engineering With Hybrid Magnonic Systems and Materials. *IEEE Trans. Quantum Eng.* **2021**, *2*, 1–36.
(16) Pezzagna, S.; Meijer, J. Quantum Computer Based on Color Centers in Diamond. *Appl. Phys. Rev.* **2021**, *8*, 011308.
(17) Yan, G. Q.; Li, S.; Yamamoto, T.; Huang, M.; Mclaughlin, N. J.; Nozaki, T.; Wang, H.; Yuasa, S.; Du, C. R. Electric-Field-Induced Coherent Control of Nitrogen-Vacancy Centers. *Phys. Rev. Appl.* **2022**, *18*, 064031.
(18) Trifunovic, L.; Pedrocchi, F. L.; Loss, D. Long-Distance Entanglement of Spin Qubits via Ferromagnet. *Phys. Rev. X* **2013**, *3*, 041023.
(19) Dolde, F.; Jakobi, I.; Naydenov, B.; Zhao, N.; Pezzagna, S.; Trautmann, C.; Meijer, J.; Neumann, P.; Jelezko, F.; Wrachtrup, J. Room-Temperature Entanglement between Single Defect Spins in Diamond. *Nat. Phys.* **2013**, *9*, 139–143.
(20) Fuchs, G. D.; Burkard, G.; Klimov, P. V.; Awschalom, D. D. A Quantum Memory Intrinsic to Single Nitrogen–Vacancy Centres in Diamond. *Nat. Phys.* **2011**, *7*, 789–793.
(21) Yao, N. Y.; Jiang, L.; Gorshkov, A. V.; Maurer, P. C.; Giedke, G.; Cirac, J. I.; Lukin, M. D. Scalable Architecture for a Room Temperature Solid-State Quantum Information Processor. *Nat. Commun.* **2012**, *3*, 800.
(22) Bradley, C. E.; Randall, J.; Abobeih, M. H.; Berrevoets, R. C.; Degen, M. J.; Bakker, M. A.; Markham, M.; Twitchen, D. J.; Taminiau, T. H. A Ten-Qubit Solid-State Spin Register with Quantum Memory up to One Minute. *Phys. Rev. X* **2019**, *9*, 031045.
(23) Zhang, J.; Hegde, S. S.; Suter, D. Efficient Implementation of a Quantum Algorithm in a Single Nitrogen-Vacancy Center of Diamond. *Phys. Rev. Lett.* **2020**, *125*, 030501.
(24) Wang, X.; Xiao, Y.; Liu, C.; Lee-Wong, E.; McLaughlin, N. J.; Wang, H.; Wu, M.; Wang, H.; Fullerton, E. E.; Du, C. R. Electrical Control of Coherent Spin Rotation of a Single-Spin Qubit. *Npj Quantum Inf.* **2020**, *6*, 78.
(25) Fuchs, G. D.; Dobrovitski, V. V.; Toyli, D. M.; Heremans, F. J.; Awschalom, D. D. Gigahertz Dynamics of a Strongly Driven Single Quantum Spin. *Science* **2009**, *326*, 1520–1522.
(26) Nozaki, T.; Shiota, Y.; Miwa, S.; Murakami, S.; Bonell, F.; Ishibashi, S.; Kubota, H.; Yakushiji, K.; Saruya, T.; Fukushima, A.; Yuasa, S.; Shinjo, T.; Suzuki, Y. Electric-Field-Induced Ferromagnetic Resonance Excitation in an Ultrathin Ferromagnetic Metal Layer. *Nat. Phys.* **2012**, *8*, 491–496.
(27) Zhu, J.; Katine, J. A.; Rowlands, G. E.; Chen, Y.-J.; Duan, Z.; Alzate, J. G.; Upadhyaya, P.; Langer, J.; Amiri, P. K.; Wang, K. L.; Krivorotov, I. N. Voltage-Induced Ferromagnetic Resonance in Magnetic Tunnel Junctions. *Phys. Rev. Lett.* **2012**, *108*, 197203.
(28) Li, S.; Huang, M.; Lu, H.; McLaughlin, N. J.; Xiao, Y.; Zhou, J.; Fullerton, E. E.; Chen, H.; Wang, H.; Du, C. R. Nanoscale Magnetic Domains in Polycrystalline $Mn_3Sn$ Films Imaged by a Scanning Single-Spin Magnetometer. *Nano Lett.* **2023**, *23*, 5326–5333.
(29) McLaughlin, N. J.; Li, S.; Brock, J. A.; Zhang, S.; Lu, H.; Huang, M.; Xiao, Y.; Zhou, J.; Tserkovnyak, Y.; Fullerton, E. E.; Wang, H.; Du, C. R. Local Control of a Single Nitrogen-Vacancy Center by Nanoscale Engineered Magnetic Domain Wall Motion. *ACS Nano* **2023**, *17*, 25689–25696.
(30) Jelezko, F.; Gaebel, T.; Popa, I.; Gruber, A.; Wrachtrup, J. Observation of Coherent Oscillations in a Single Electron Spin. *Phys. Rev. Lett.* **2004**, *92*, 076401.
(31) Solyom, A.; Flansberry, Z.; Tschudin, M. A.; Leitao, N.; Pioro-Ladrière, M.; Sankey, J. C.; Childress, L. I. Probing a Spin Transfer Controlled Magnetic Nanowire with a Single Nitrogen-Vacancy Spin in Bulk Diamond. *Nano Lett.* **2018**, *18*, 6494–6499.

(32) Chen, Y.-J.; Lee, H. K.; Verba, R.; Katine, J. A.; Barsukov, I.; Tiberkevich, V.; Xiao, J. Q.; Slavin, A. N.; Krivorotov, I. N. Parametric Resonance of Magnetization Excited by Electric Field. *Nano Lett.* **2017**, *17*, 572–577.

(33) Lee-Wong, E.; Xue, R.; Ye, F.; Kreisel, A.; Van Der Sar, T.; Yacoby, A.; Du, C. R. Nanoscale Detection of Magnon Excitations with Variable Wavevectors Through a Quantum Spin Sensor. *Nano Lett.* **2020**, *20*, 3284–3290.

(34) Labanowski, D.; Bhallamudi, V. P.; Guo, Q.; Purser, C. M.; McCullian, B. A.; Hammel, P. C.; Salahuddin, S. Voltage-Driven, Local, and Efficient Excitation of Nitrogen-Vacancy Centers in Diamond. *Sci. Adv.* **2018**, *4*, eaat6574.

(35) Gottscholl, A.; Diez, M.; Soltamov, V.; Kasper, C.; Sperlich, A.; Kianinia, M.; Bradac, C.; Aharonovich, I.; Dyakonov, V. Room Temperature Coherent Control of Spin Defects in Hexagonal Boron Nitride. *Sci. Adv.* **2021**, *7*, eabf3630.